\documentstyle[prl,preprint,aps,epsf]{revtex}
\begin{document}
\draft
\preprint{GTP-97-04}
\title{ The Second Virial Coeffecient of Spin-1/2 Anyon System}
\author
{$^{a}$Sahng-kyoon Yoo and $^{b}$D. K. Park}
\address
{$^{a}$Department of Physics, Seonam University, Namwon 590-711, Korea \\
 $^{b}$Department of Physics, Kyungnam University, Masan 631-701, Korea}
\date{\today}
\maketitle
\begin{abstract}
We calculate the second virial coefficient of spin-1/2 anyon gas in the 
various values of the self-adjoint extension parameter by
incorporating the self-adjoint extension method into the Green's function
formalism. Especially, the completely different cusp- and 
discontinuity-structures
from the result of previous literature are obtained 
when the self-adjoint extension
parameter goes to infinity. This is originated from the different condition
for the occurrence of irregular states.
\end{abstract}
\newpage

Since the Kronig-Penny model\cite{kronig31} described successfully the band
structure of energy spectrum in the solid-state physics, the point interaction
problem has been applied to the various branches of physics for a long time.
Recent application of the point interaction in the theoretical physics seems
to be concentrated upon the subjects related to the quantum mechanical
renormalization\cite{adhikari95,phillips97} and anyonic theories.\cite{chen89}

\indent The frequent use of the point interaction in the quantum mechanical
renormalization is mainly due to its advantage of permitting the derivation
of the exact solution. Hence, the comparision of an exact solution with a
perturbative solution, which can be obtained by solving the Lippmann-Schwinger
equation iteratively, allows us to understand the subtleties of renormalization
scheme encountered frequently in the quantum field theories. It may be also
very helpful to understand the highly non-trivial concepts like dimensional
transmutation\cite{thron79} and anomalies\cite{jackiw85}.

\indent In spin-1/2 Aharonov-Bohm problem\cite{gerbert89,hagen91},
which is directly related to the fermion-based anyonic theory,
the two-dimensional point interaction is realized as a Zeeman interaction
of the spin with a magnetic flux tube.
The two different approaches, renormalization and self-adjoint
extension\cite{albeverio88}, for the quantum mechanical point interaction
and the coincidence of their results are presented by Jackiw\cite{jackiw91}.
In Ref. \cite{park96} one of us analyzed the incorporation
of the self-adjoint extension into the Green's function formalism
by solving the Lippmann-Schwinger equation nonperturbatively.

\indent One of the way to investigate the statistical properties of the 
spin-1/2 anyon system is to evaluate the second virial coefficient as a 
function of a statistical parameter.
The second virial coefficient of the
Boson-based anyonic system is firstly calculated by Arovas {\it et al}.\cite{arovas85}.
The most interesting feature of their result is that
the virial coefficient interpolates between the Bose and Fermi values with
a periodic dependence on the flux $\alpha$ carried by anyon. Subsequently,
the second virial coefficient for a system of spin-1/2 anyon gas is computed
by Blum {\it et al}.\cite{blum90} by using the condition of the irregular solution
\begin{eqnarray}
& \mid m+ \alpha \mid < 1 , & \nonumber \\
& \mid m \mid + \mid m+ \alpha \mid = - \alpha s, &
\label{condition1}
\end{eqnarray}
where $m$ is the angular momentum quantum number and $s$ is $\pm 1$ for
spin up and spin down, respectively. Their striking result is that there
exist discontinuities in the virial coefficient at all even, nonzero values
of $\alpha$.

\indent In this Letter we compute the second virial coefficient for a
spin-1/2 anyon gas by incorporating the self-adjoint extension method into
the Green's function formalism. In this case, it is well-known that the
arbitrary combination of the regular and irregular solution, which is
characterized by the self-adjoint extension parameter, is derived when
$\mid m+ \alpha \mid < 1$. 
We expect the absence of the latter condition
in Eq.(\ref{condition1}) for the appearence of the irregular solution
in the self-adjoint extension method 
does yield completely different cusp- and
discontinuity-structure in the virial coefficient from the known in the
previous literatures.
The evaluation is performed at various self-adjoint extension parameter
$\lambda_m$.  
In particular, we find  
new discontinuities at all non-zero integer values of flux
when $\lambda_m \rightarrow \infty$.
When, however, $\lambda_m$ becomes finite, the discontinuities turn into cusps.
This might be an existence of different kind of phase transitions from the
previous cases.

\indent In general, the second virial coefficient can be calculated by making 
an energy spectrum discrete. Therefore, we introduce the simple harmonic 
oscillator potential in the Hamitonian 
\begin{equation}
H = H_0 + v \delta ( {\bf r} ),
\end{equation}
where
\begin{equation}
H_0 = \frac{1}{2M} ( {\bf p} - e {\bf A} )^2 +
\frac{M \omega^2}{2} {\bf r}^2.
\end{equation}
The energy-dependent Green's function $\hat{G} [{\bf r_1}, {\bf r_2}: E]$
for $H$ can be derived by solving the Lippmann-Schwinger equation
nonperturbatively. Following Ref. \cite{park96} one can arrive at
\begin{equation}
\hat{A} [{\bf r_2}, {\bf r_1}:E] = \frac{
    \hat{G_0}  [ {\bf r_2}, {\bf \epsilon_1} : E]
    \hat{G_0}  [ {\bf \epsilon_2}, {\bf r_1} : E]}
    { \frac{1}{v} + \lim_{ {\bf \epsilon_2} \rightarrow {\bf \epsilon_1}^+ }
    \hat{G_0}  [ {\bf \epsilon_2}, {\bf \epsilon_1} : E]},
\end{equation}
where
\begin{equation}
\hat{A} [ {\bf r_2}, {\bf r_1}: E] \equiv
\hat{G} [ {\bf r_2}, {\bf r_1}: E] -
\hat{G_0} [ {\bf r_2}, {\bf r_1}: E]
\end{equation}
and $\hat{G_0} [ {\bf r_2}, {\bf r_1}: E] $ is energy-dependent Green's
function for $H_0$ \cite{peak69}. Since
$\hat{G} [ {\bf r_2}, {\bf r_1}: E] =  \sum_{n} \phi_n^* ({\bf r_2})
\phi_n ({\bf r_1}) / (E + E_n)$,  $\hat{G}$ and $\phi$ must
obey the same boundary condition, which is given by self-adjoint extension
method at the origin \cite{15}. 
In Ref. \cite{park96} bound state energy is derived by imposing the
boundary condition to $\hat{A} [ {\bf r_2}, {\bf r_1} :E]$ instead of
$\hat{G} [ {\bf r_2}, {\bf r_1} :E]$. In this case the bound state 
spectrum can be 
obtained by solving the equation
\begin{equation}
- \frac{1}{\lambda_m} = (M \omega)^{ \mid m+ \alpha \mid}
\frac{ \Gamma (1- \mid m+ \alpha \mid )}{ \Gamma (1+ \mid m+ \alpha \mid )}
\frac{ \Gamma \left( \frac{1+ \mid m+ \alpha \mid + E/ \omega }{2} \right)}
     { \Gamma \left( \frac{1- \mid m+ \alpha \mid + E/ \omega }{2} \right)},
\label{en1}
\end{equation}
when $\mid m+ \alpha \mid < 1$.
Here, $\lambda_m$ is the self-adjoint extension
parameter. 
If $\hat{G_0} [ {\bf r_2}, {\bf r_1} :E]$
does not have any pole, the imposition of the boundary condition to 
$\hat{A} [ {\bf r_2}, {\bf r_1} :E]$ is physically relevant because the
most contribution to $\hat{G} [ {\bf r_2}, {\bf r_1} :E]$ is
$\hat{A} [ {\bf r_2}, {\bf r_1} :E]$ at bound state energies. If, however,
$\hat{G_0} [ {\bf r_2}, {\bf r_1} :E]$ does have poles, we can not conclude
{\it a priori} the relevance of the above procedure. In order to get the
credibility in this case, the poles in $\hat{G_0} [ {\bf r_2}, {\bf r_1} :E]$ 
must not contributed to the
poles in $\hat{G} [ {\bf r_2}, {\bf r_1} :E]$ at the final stage. We confirmed
by explicit calculation that this is indeed the case in our model.
Furthermore, imposing the boundary condition to $\hat{G} [ {\bf r_2}, {\bf r_1} :E]$,
one arrives at the same result.  

When $\mid m+ \alpha \mid > 1$, no irregular solution occurs, and
therefore, the bound state energies are given by
\begin{equation}
\epsilon_{n,m} = (2n +1 + \mid m + \alpha \mid ) \omega,
\label{bound1}
\end{equation}
where $n$ is a non-negative integer and $m$ is an integer. We first consider
the two special cases in Eq.(\ref{en1}): $\lambda_m =0$ and $\lambda_m = \infty$. In both
cases, we can obtain the analytic solutions for the bound state
spectrum. For the former
case, the bound state energies coincide with Eq.(\ref{bound1}), which
means that there is no irregular states even at $\mid m+ \alpha \mid < 1$.
(hard-core repulsion potential exists.) For the latter case, however, the
bound state energies are given by
\begin{equation}
\epsilon_{n,m} = (2n +1 - \mid m+ \alpha \mid ) \omega.
\end{equation}
This is exactly the same as the case of Blum {\it et al}.\cite{blum90}. 
The only difference is that there is no 
$\mid m \mid + \mid m+ \alpha \mid = - \alpha s $ 
condition for the occurrence of the irregular solution 
in our self-adjoint extension method.
We plot the energy $\epsilon_{0,0}$ for
$\lambda_m = \infty$ as a
function of $\alpha$ in Fig. \ref{figure1}. In contrast to Ref. 
\cite{blum90}, there are two discontinuities at $\alpha =-1$ and $+1$,
respectively, and one cusp at $\alpha = 0$. 
This will probably imply the different behavior of second
virial coefficient from that of Blum {\it et al}.

\indent When $0 < \lambda_m < \infty$, the bound sate spectrum for 
$\mid m+ \alpha \mid < 1$ can be found by graphical analysis. The splitting
between the adjascent levels is not $2 \omega$ unlike the two above-mentioned
cases, 
but depends on $\lambda_m$ and $\alpha$. Using the energy eigenvalues
obtained numerically, we now calculate the second virial coefficient 
\begin{equation}
B_2 (T) - B_2^0 (T) = -2 \lambda_T^2 
   \sum_{n,m} \left[ e^{-\beta \epsilon_{n,m}} - 
                     e^{- \beta \epsilon_{n,m} (\alpha =0)} \right],
\end{equation}
where $B_2^0 (T)$ is the second virial coefficient of free Fermion (Boson)
which is given by $\frac{1}{4} \lambda_T^2$ ($- \frac{1}{4} \lambda_T^2 )$,
$\lambda_T$ is the thermal de Broglie wavelength and $\beta = 1 / kT$.

\indent First let us consider the Boson-based case. We perform the summation only
over the even $m$, and obtain
\begin{eqnarray}
B_2^B (T) - B_2^{0,B} (T) & = & -2 \lambda_T^2 \nonumber \\ \nonumber \\
& \times & \left[
   \begin{array}{llll} 
     \frac{e^{-\beta \omega}}{2 \sinh \beta \omega}
      [ \cosh \beta \delta \omega -1]  & \\ \\
     + \sum_{n, \mid m+ \alpha \mid <1} \left[ e^{-\beta \epsilon_{n, -N}}
       - e^{-\beta \epsilon_{n, -N} (\alpha=0)} \right],
       & \mbox{ $N =$ even}, \\ \\
     \frac{e^{-\beta \omega}}{2 \sinh \beta \omega}
     [ \cosh \beta (1-\delta) \omega -\cosh \beta \omega] & \\ \\
     + \sum_{n, \mid m+ \alpha \mid <1} \left[ e^{-\beta \epsilon_{n, -N-1}}
       - e^{-\beta \epsilon_{n, -N-1} (\alpha=0)} \right],
       & \mbox{ $N =$ odd}, \\
     \end{array} \right.
\end{eqnarray}
where $\alpha = N + \delta$, $N$ is an integer and $0 < \delta < 1$. In the
$\omega \rightarrow 0$ limit, the result for $B_2^B (T) / \lambda_T^2$
at different self-adjoint extension
parameters are shown in Fig. \ref{figure2}. 
The simple numerical calculations
are needed to perform the summations when $\mid m+ \alpha \mid < 1$.
We add the $e^{-\beta \epsilon_{n,m}}$ factors up to $n=5$, for which the
errors are less than 1 \%. When $\lambda_m = 0$, the result is exactly same as 
that of Arovas {\it et al}.\cite{arovas85} When $\lambda_m$ begins to be finite, 
the cusps at the Fermion points begin to appear and become more deeper with
increasing $\lambda_m$. This fact agrees with the recent 
calculation of second virial coefficient of anyons without hardcore
\cite{kim97}. 
The appearence of cusps at the Fermion point is interpreted in Ref.
\cite{loss91} as follows: 
as the
possibility of the overlap between particles is introduced, the particles
becomes more Boson-like. 
The cusps at the Bose points still remain, at variance with the result of 
Ref. \cite{kim97} which is unbelievable on account of the abrupt 
appearence of the inflection point at 
$ \epsilon = 10 $.  
As the self-adjoint extension parameters increase, the values
at all points except the Bose point become smaller than the hard-core values.
This can be considered as follows.
The introduction of the irregular states at the origin increases the 
possibility of overlap between particles, so that it makes
the system compressed.
Increasing the portion of the irregular solution in wave function, 
the augmentation of the self-adjoint extension parameter decreases the second
virial coefficient.

\indent The Fermion-based case can be calculated by summing odd $m$'s. We 
consider only the unpolarized Fermion system, which is obtained by averaging
over the four spin states (triplet + singlet). The virial coefficient
$B_2^F (T) / \lambda_T^2$ in the $\omega \rightarrow 0$ is plotted in 
Fig. \ref{figure3} for $\lambda_m =0, 1/5, 1$ and $\infty$. 
When $\lambda_m =0$,
the result without irregular solution 
in Ref. \cite{blum90} is recovered. 
But for finite $\lambda_m$, the value at Bose point jumps down
suddenly to the negative value and the second
virial coefficient decreases with increasing $\lambda_m$ as in the Boson-based case. As 
$\lambda_m \rightarrow \infty$ (the case of Blum {\it et al}\cite{blum90}), 
there exist two
kinds of discontinuity at both Boson and Fermion points, which shows the 
different behavior from Ref. \cite{blum90}. 
This means that it is some
different phase transitions from those of Blum {\it et al}. The origin of this
difference comes from the different condition for the occurence of irregular
states. As $\lambda_m$ decreases, discontinuities turn into cusps at integer
points and as $\lambda_m \rightarrow 0$, the result is reduced to the case in
which the irregular solution is ignored. 

\indent In conclusion, using the self-adjoint extension method and imposing
the boundary conditions at the origin, 
we calculate the second virial coefficients for the Boson-based and 
Fermion-based 
anyon systems. Both cases show
the decreasing behaviors when the self-adjoint extension parameter $\lambda_m$
increases. This tells us that the values of parameters determine the magnitude
of repulsion between two particles. Although, for the Boson-based case, 
the decreasing behavior also shows up in Ref. \cite{kim97}, the cusp structure
at Bose point is completely different.
It is also shown that the Fermion-based
case has a different kind of discontinuity from that of Blum {\it et al}.
\cite{blum90}, which is due to the absence of the second condtition
in Eq. (\ref{condition1}).
This might exhibit another kind of phase transition.

\begin{figure}

\caption{ The bound state energies $\epsilon_{0,0}$ as a function of
   $\alpha$ for $\lambda_m = \infty$ case. }
\label{figure1}
\end{figure}

\begin{figure}

\caption{ $B_2^B (T) / \lambda_T^2 $ for Boson-based anyons at various
   self-adjoint extension parameters $\lambda_m$. }
\label{figure2}   
\end{figure}

\begin{figure}

\caption{ $B_2^F (T) / \lambda_T^2 $ for unpolarized Fermion-based anyons 
at various self-adjoint extension parameters $\lambda_m$. }
\label{figure3}   
\end{figure}
\newpage
\epsfysize=25cm \epsfbox{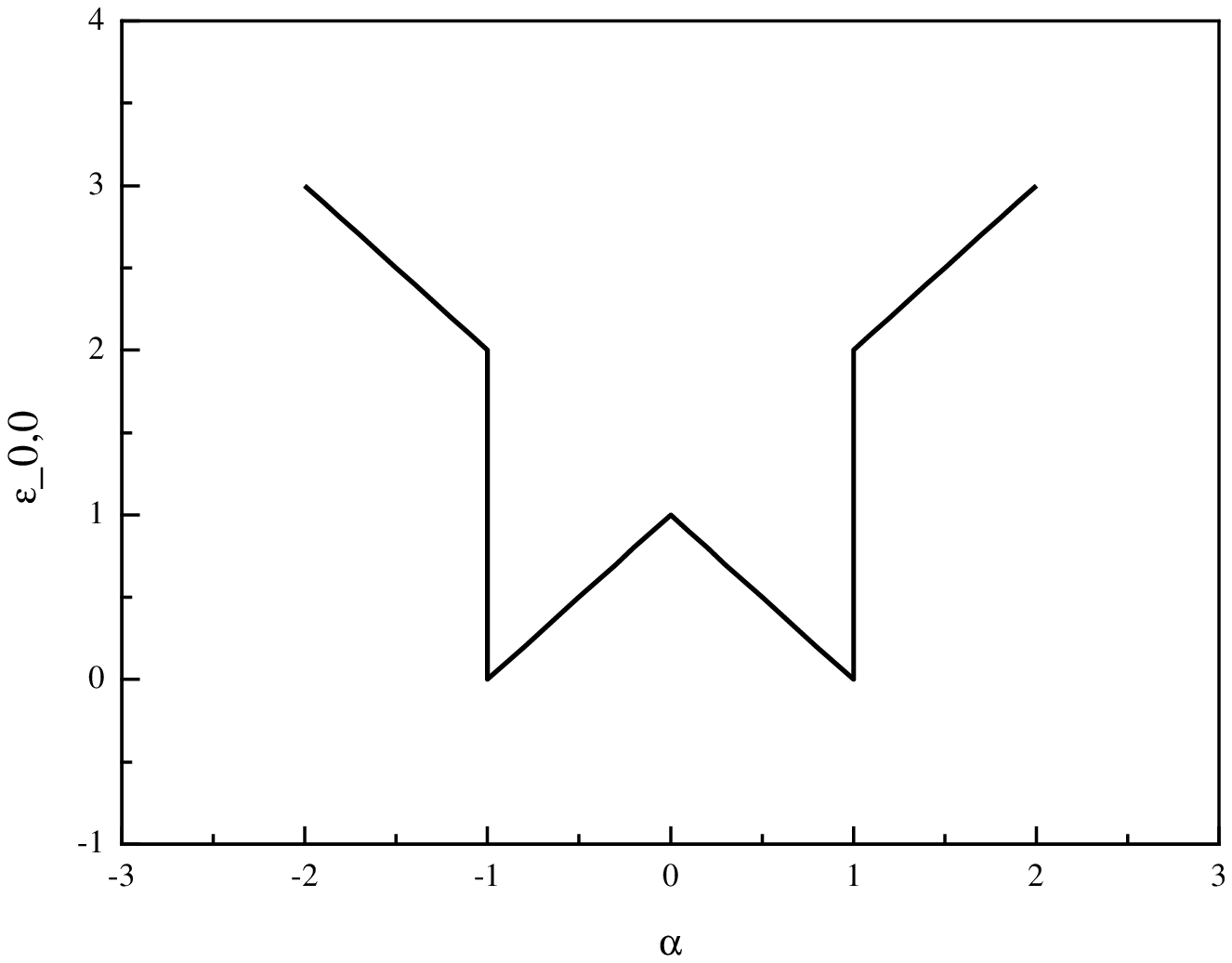}
\newpage
\epsfysize=25cm \epsfbox{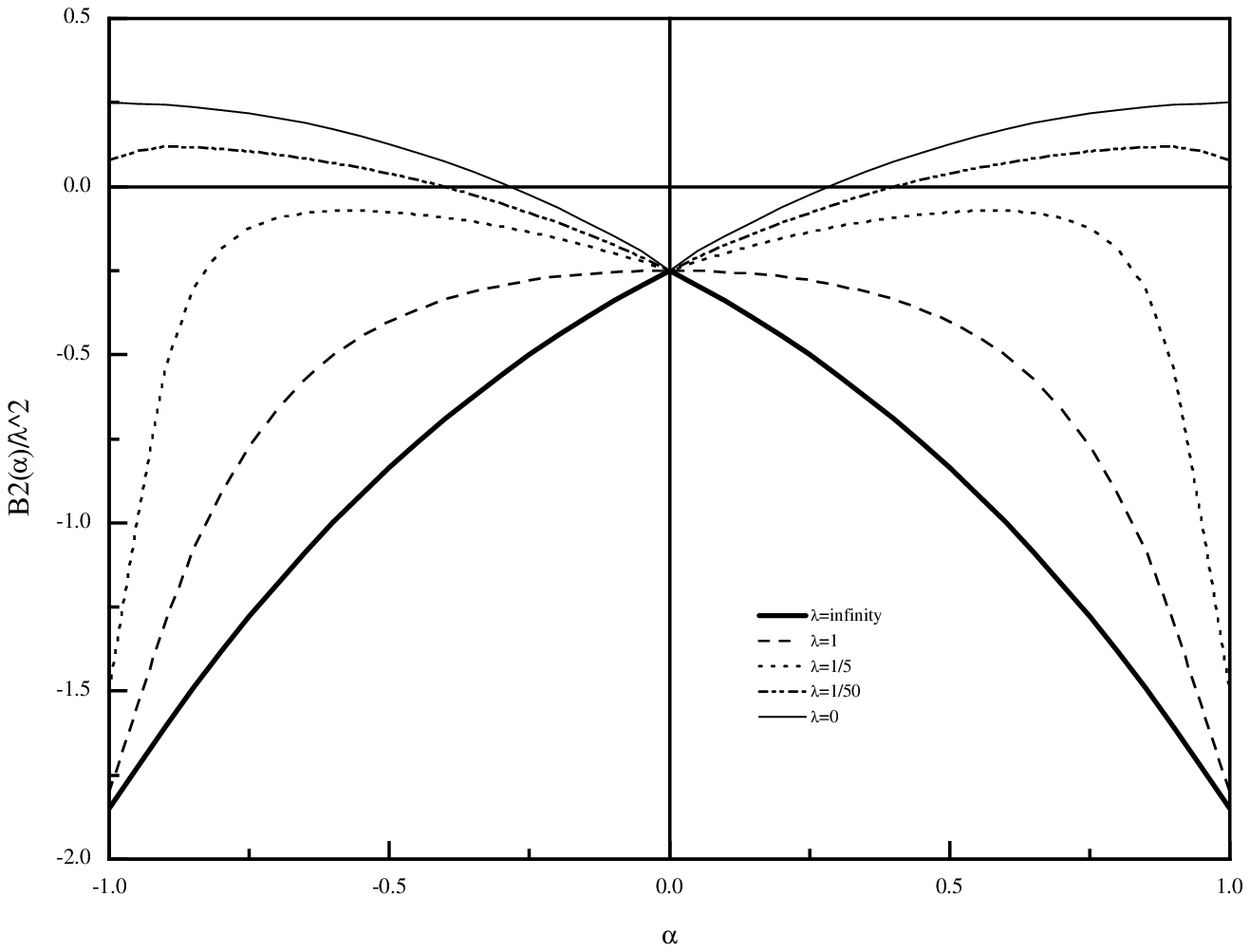}
\newpage
\epsfysize=25cm \epsfbox{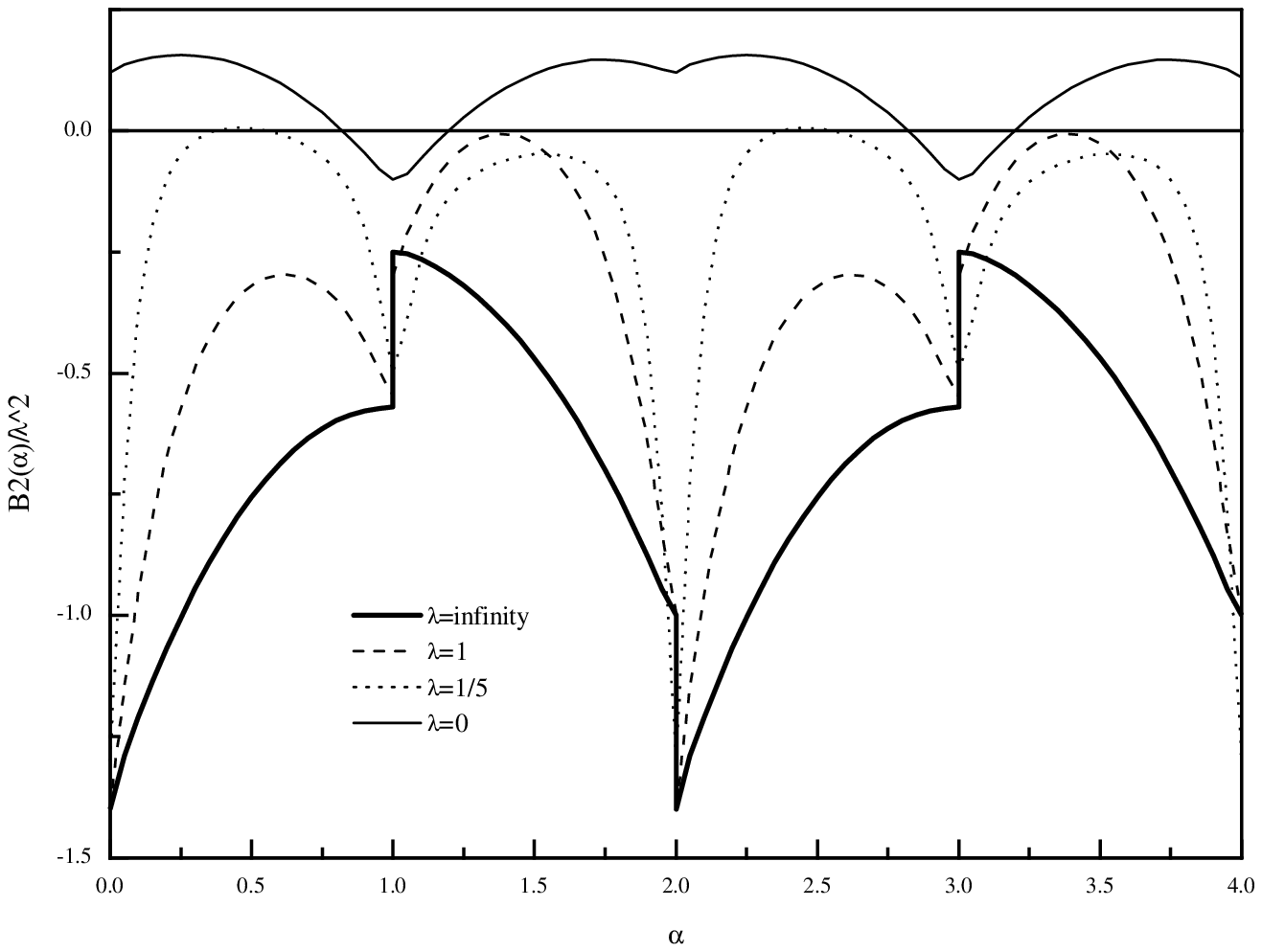}

\begin{references}
\bibitem{15} The explicit boundary condition is given in Eq.(52) of Ref. \cite{park96}.
\end{references}
\end{document}